\documentclass[12pt]{article}
\usepackage{amsfonts}
\usepackage{amsmath}
\usepackage{amssymb}
\usepackage{graphicx}
\usepackage{color}
\usepackage[all, knot]{xy}
\usepackage{tikz}

\usepackage{ulem}

\usepackage[utf8]{inputenc}
\usepackage{epstopdf}
\usepackage[footnotesize]{caption}
\usepackage{amsthm}
\usepackage{enumitem}
\usepackage{mathrsfs}

\usepackage[margin=3cm]{geometry}

\def \be {\begin{equation}}
\def \ee {\end{equation}}
\def \bea {\begin{eqnarray}}
\def \eea {\end{eqnarray}}
\def \nn {\nonumber}

\def \rr {\raise.35ex\hbox{\small $\prime$}\kern-.17em{\mbox{\large $\imath$}}}

\def \dels {\partial\kern-.6em /\kern.1em}
\def \As {{A\kern-.5em / \kern.5em}}
\def \Ds {D\kern-.7em / \kern.5em}

\def \ks {k\kern-.5em /}
\def \ls {l\kern-.5em /}

\newcommand{\ci}[1]{}
\newcommand{\ba}{\begin{eqnarray}}
\newcommand{\ea}{\end{eqnarray}}
\newcommand{\bal}{\begin{align}}
\newcommand{\eal}{\end{align}}
\newcommand{\bay}[1]{\left(\begin{array}{#1}}
\newcommand{\eay}{\end{array}\right)}

\newcommand{\hide}[1]{}

\DeclareMathOperator{\sech}{sech}

\newlist{axioms}{enumerate}{2}
\setlist[axioms,1]{label=\textbf{A\arabic{axiomsi}.}, ref=A\arabic{axiomsi}}
\setlist[axioms,2]{label=\textbf{A\arabic{axiomsi}\rlap{\myEnumCounter{axiomsii}}.},%
                   ref=A\arabic{axiomsi}\myEnumCounter{axiomsii},%
                   align=parleft,%
                   leftmargin=0em,%
                   itemsep=1.4ex,%
                   before={\stepcounter{axiomsi}}}

  \usetikzlibrary{decorations.markings}

\begin{document}

\begin{titlepage}
\begin{center}

\textbf{\LARGE
Modular Average and Weyl Anomaly in\\ 
Two-Dimensional Schwarzian Theory
\vskip.3cm
}
\vskip .5in
{\large
Xing Huang$^{a,b,c}$ \footnote{e-mail address: xingavatar@gmail.com} and
Chen-Te Ma$^{d,e,f,g,h,i}$ \footnote{e-mail address: yefgst@gmail.com}
\\
\vskip 1mm
}
{\sl
$^a$
Institute of Modern Physics, Northwest University, Xi'an 710069, China.
\\
$^b$
NSFC-SPTP Peng Huanwu Center for Fundamental Theory, Xi'an 710127, China.
\\
$^c$
Shaanxi Key Laboratory for Theoretical Physics Frontiers, Xi'an 710069, China. 
\\
$^d$ 
Department of Physics and Astronomy, Iowa State University, Ames, Iowa 50011, US.
\\
$^e$
Asia Pacific Center for Theoretical Physics,\\
Pohang University of Science and Technology, 
Pohang 37673, Gyeongsangbuk-do, South Korea. 
\\
$^f$
Guangdong Provincial Key Laboratory of Nuclear Science,\\
 Institute of Quantum Matter,
South China Normal University, Guangzhou 510006, Guangdong, China. 
\\ 
$^g$ 
School of Physics and Telecommunication Engineering,\\
South China Normal University, Guangzhou 510006, Guangdong, China. 
\\ 
$^h$
Guangdong-Hong Kong Joint Laboratory of Quantum Matter,\\
 Southern Nuclear Science Computing Center, 
South China Normal University, 
Guangzhou 510006, Guangdong, China. 
\\
$^i$ 
The Laboratory for Quantum Gravity and Strings,\\
 Department of Mathematics and Applied Mathematics, 
University of Cape Town,
 Private Bag, Rondebosch 7700, South Africa.
}\\
\vskip 1mm
\vspace{40pt}
\end{center}

\newpage
\begin{abstract} 
The gauge formulation of Einstein gravity in AdS$_3$ background leads to a boundary theory that breaks modular symmetry and loses the covariant form. 
We examine the Weyl anomaly for the cylinder and torus manifolds. 
The divergent term is the same as the Liouville theory when transforming from the cylinder to the sphere. 
The general Weyl transformation on the torus also reproduces the Liouville theory.  
The Weyl transformation introduces an additional boundary term for reproducing the Liouville theory, which allows the use of CFT techniques to analyze the theory. 
The torus partition function in this boundary theory is one-loop exact, and an analytical solution to disjoint two-interval Rényi-2 mutual information can be obtained. 
We also discuss a first-order phase transition for the separation length of two intervals, which occurs at the classical level but is smoothed out by non-perturbative effects captured by averaging over a modular group in the boundary theory. 
%The boundary conditions for the A-cycle and the B-cycle torus are different.  
%Therefore, it is not a conventional CFT, and the Lagrangian loses the covariant form. 
%The mutual information should decay for a long separation of two intervals. 
%The averaging over a modular group respects the expectation. 
%Because not all saddle points provide the decay behavior, our result should be non-trivial evidence to support the modular averaging as a non-perturbative completion.    
%Introducing an ensemble to a boundary theory is a simple implementation of the non-perturbative effect. 
%The averaging of a modular group in ${\cal N}=4$ SYM theory in the small gravitational constant limit is equivalent to the strong-coupling limit of a planar diagram. 
%The exact bulk dual of ensemble CFT is necessary to have a non-perturbative completion of gravity theory. 
%Our study is a toy study of a page curve. 
%It is easy to implement the non-perturbative effect by introducing an ensemble to a boundary theory. 
\end{abstract}
\end{titlepage}

\section{Introduction}
\label{sec:1}
\noindent 
The concept of entanglement is a fundamental one in Quantum Mechanics, and it is essential to understand how it is related to the geometry of space-time to gain insights into the nature of gravity. 
One of the most intriguing phenomena in this regard is the {\it black hole information paradox}, which arises from the apparent loss of information in a black hole. 
The {\it semi-classical} calculation showed that an isolated black hole allows physical information to disappear. 
However, the principle of Quantum Mechanics implies unitarity in a closed system, which then means the contradiction between General Generality and Quantum Mechanics. 
The paradox has been the subject of intense research in theoretical physics for decades and recent progress in understanding the role of quantum entanglement and holography or Page curve has provided a potential pathway to a solution. 
The concept of holography, which suggests that the information of a black hole is encoded on its horizon, also plays a crucial role in this research direction. 
It is hoped that a better understanding of the paradox can provide insight into the nature of quantum gravity and the ultimate fate of information in the universe. 
The Page curve describes the amount of information (or entanglement entropy) between a black hole and the radiation, which forms a closed system with the black hole and the incoming waves. 
The curve implies that entanglement entropy first increases over time and then decreases to zero, with the semi-classical calculation losing the decreasing behavior. 
However, considering the sum of saddle points (or the replica wormhole solutions) in the gravity path integration generates the Page curve \cite{Penington:2019kki,Almheiri:2019qdq}, and the increasing and decreasing behaviors do not correspond to the same saddle-point solution \cite{Penington:2019kki,Almheiri:2019qdq}. 
Therefore, the information is possibly recovered after taking into account the non-perturbative effect properly. 
\\ 

\noindent 
The wormhole induces a spectrum that needs to introduce an {\it ensemble} to boundary field theory \cite{Cotler:2021cqa}.  
One discussed the ensemble on a connected bulk manifold with the disconnected boundary. 
The ensemble averaging does {\it not} affect some fixed energy observables for the Einstein gravity theory (the higher-dimensional case is also discussed in Ref. \cite{Schlenker:2022dyo}). 
This fact arises from two assumptions \cite{Schlenker:2022dyo}: 
\begin{itemize}
\item The fixed energy observables are not relevant to an ensemble averaging. 
The black hole physics should be chaotic and have an ensemble averaging. 
\item The Hamiltonian $H_N$ does {\it not} have a large $N$ limit for describing a single black hole \cite{Schlenker:2022dyo}. 
\end{itemize}
Combining two assumptions implies the independent description of the random matrix for $H_{N}$ for each $N$ \cite{Schlenker:2022dyo}, which means that the self-averaging of the ensemble is necessary for obtaining the boundary theory.  
In other words, the description is not a smooth function for $N$.   
In String Theory, the  Anti-de Sitter/Conformal Field Theory (AdS/CFT) correspondence has a smooth description from $N$. 
The boundary dual of Einstein gravity theory is not the conventional AdS/CFT correspondence \cite{Schlenker:2022dyo}. 
However, ensemble averaging is convenient for analyzing the non-perturbative contribution. 
The {\it modular averaging} in Einstein gravity is equivalent to an {\it ensemble average} of the large central charge CFTs \cite{Chandra:2022bqq}, and it can be used to compute non-perturbative contributions to the boundary theory. 
Hence we are interested in the analytical implementation of the averaging of a modular group in Einstein gravity theory. 
\\

\noindent
The AdS$_3$ Einstein gravity theory \cite{Mikhaylov:2017ngi,Collier:2023fwi} is topological at the classical level, and all physical degrees are on the boundary, which allows for deriving the boundary theory. 
On the torus, the boundary theory has a similar form to the Schwarzian theory (called 2d Schwarzian theory) \cite{Cotler:2018zff}, which does {\it not} respect modular symmetry. 
The breaking of the modular symmetry can happen when the conformal fields on the torus do not follow the periodic boundary condition for each cycle. 
The torus partition functions for the A- and B-cycles are the same, only for the conventional CFT.  
However, the bulk theory guarantees that the boundary theory has the Lorentz symmetry and the Weyl invariance at the classical level. 
Hence the boundary theory is conformal but loses the torus' modular symmetry.  
The path integration over bulk manifolds for the asymptotic AdS$_3$ boundary is equivalent to summing over a modular group \cite{Maloney:2007ud,Manschot:2007ha}, which should provide the instanton effect to include all bulk geometries. 
Therefore, the sum over a modular group \cite{Maloney:2007ud,Manschot:2007ha} should provide a non-perturbative completion with a manifest modular symmetry. 
The modular summation leads to the negative density of states. 
The averaging ensemble cannot provide the unitary CFT result, which is also supported by the CFT computation \cite{Mukhametzhanov:2019pzy,Ganguly:2019ksp,Kusuki:2019gjs,Benjamin:2019stq,Maxfield:2019hdt}. 
The study of the AdS$_3$/CFT$_2$ correspondence has led to the discovery of new dualities and relations between different areas of physics, such as the connection between quantum chaos and black hole. 
An understanding of the boundary theory and its symmetries is crucial for exploring these connections and for further developments in the field. 
For example, the exploration of Goldstone effective theories from SL(2, $\mathbb{R}$) symmetry group \cite{Anninos:2021ydw}.
\\

\noindent
In this paper, we explore the connection between the Weyl transformation and the Liouville theory and apply the results to calculate the Rényi-2 mutual information for disjoint two intervals \cite{Lunin:2000yv,Headrick:2010zt,Faulkner:2013yia} in the boundary theory. 
When the Weyl anomaly is given by the Liouville theory, one can apply the CFT result to AdS$_3$ Einstein gravity theory. 
One calculated the Rényi entropy for a single interval by generalizing the torus to the $n$-sheet torus \cite{Huang:2019nfm,Huang:2020tjl}. 
The result is one-loop exact (all effects disappear beyond one loop). 
Therefore, Rényi-2 mutual information for two intervals is calculable and has an analytical expression. 
Rényi entropy is computed in Lorentzian pure Jackiw-Teitelboim gravity with a Hilbert space that can be factored not by a spatial direction \cite{Jafferis:2019wkd}. 
We calculate Rényi-2 mutual information for disjoint two-intervals in the boundary theory. 
The Rényi-2 mutual information shows a first-order phase transition in the absence of non-perturbative effects, and a smooth curve connecting two phases when such effects are considered (implemented by the modular average). 
We also discuss the decay behavior of the mutual information for large separation lengths. 
To summarize our results:
\begin{itemize}
\item 
The Liouville theory is induced by this anomaly using the covariance in CFT$_2$. 
The choice of the background leads to the non-covariant Lagrangian formulation under coordinate transformations.  
We discuss the Weyl anomaly in the boundary theory because it loses the covariance. 
From the cylinder to the sphere, the divergent term is the same as the Liouville theory. 
The general Weyl transformation on the torus manifold also gives the Liouville theory. 
The Liouville term needs the additional boundary term, not the same as in the claim of Ref. \cite{Cotler:2018zff}. 
Hence one can apply the CFT result to the Rényi-2 mutual information. 
\item We obtain the analytical solution to the Rényi-2 mutual information for two disjoint intervals. 
The first-order phase transition for the separation length of two intervals happens if one only concerns the classical gravity. 
Our result shows that the non-perturbation effect kills the phase transition. 
\item Although not all saddle solutions provide the decay behavior to the mutual information for a long separation (like $B$ cycle), the averaging over a modular group shows the expected decay.  
\end{itemize}

\noindent
The organization of this paper is as follows: 
We discuss the Weyl anomaly and its relationship with the Weyl transformation on the cylinder and torus manifolds in Sec.~\ref{sec:2}. 
We then describe the implementation of modular averaging to 2d Schwarzian theory in Sec.~\ref{sec:3}. 
The result of Rényi-2 mutual information for two disjoint intervals is in Sec.~\ref{sec:4}. 
Finally, we discuss the results presented in the previous section and provide a conclusion in Sec.~\ref{sec:5}. 

\section{Weyl Anomaly}
\label{sec:2} 
\noindent 
We first review the gauge formulation of AdS$_3$ Einstein gravity theory and the boundary theory \cite{Cotler:2018zff}. 
The boundary theory on the torus does not have modular symmetry. 
The Lagrangian is also not covariant under coordinate transformations. 
Because the boundary theory is not a conventional CFT, we explicitly check the Weyl transformation on the cylinder and torus. 
The result is consistent with the Liouville theory without concerning the one-dimensional boundary term.    

\subsection{Review of Gauge Formulation} 
\noindent 
The AdS$_3$ geometry is 
\bea
ds^2=-(r^2+1)dt^2+\frac{dr^2}{r^2+1}+r^2d\theta^2,
\eea
in which the ranges of coordinates are defined by:
\bea
-\infty< t<\infty; \qquad 0<r<\infty; \qquad 0<\theta\le 2\pi \,.
\eea

\noindent 
The Lagrangian description for the gauge formulation of AdS$_3$ Einstein gravity theory is
\bea
&&S_{\mathrm{G}}
\nn\\
&=&\frac{k}{2\pi}\int d^3x\ \mathrm{Tr}\bigg(A_tF_{r\theta}-\frac{1}{2}\big(A_r\partial_tA_{\theta}-A_{\theta}\partial_tA_r\big)\bigg)
\nn\\
&&-\frac{k}{2\pi}\int d^3x\ \mathrm{Tr}\bigg(\bar{A}_t\bar{F}_{r\theta}-\frac{1}{2}\big(\bar{A}_r\partial_t\bar{A}_{\theta}-\bar{A}_{\theta}\partial_t\bar{A}_r\big)\bigg)
\nn\\
&&-\frac{k}{4\pi}\int dtd\theta\ \mathrm{Tr}\bigg(\frac{E_t^+}{E_{\theta}^+}A_{\theta}^2\bigg)
\nn\\
&&+\frac{k}{4\pi}\int dtd\theta\ \mathrm{Tr}\bigg(\frac{E_t^-}{E_{\theta}^-}\bar{A}_{\theta}^2\bigg), 
\eea
where 
\bea
F_{\mu\nu}^a\equiv\partial_{\mu}A_{\nu}^a-\partial_{\nu}A_{\mu}^a+\lbrack A_{\mu}, A_{\nu}\rbrack^a. 
\eea 
We just replace $A$ with $\bar{A}$ in another SL(2) field strength $\bar{F}_{\mu\nu}$. 
The $F_{r\theta}$ and $\bar{F}_{r\theta}$ are the $r$-$\theta$ components of the field strengths. 
We label the index of Lie algebra by $a$. 
The gauge fields are relevant to the dreibeine $e_{\mu}\equiv e_{\mu}{}^aJ_a$ and spin connection 
$\omega_{\mu}\equiv\omega_{\mu}{}^aJ_a$:
\bea
A_{\mu}\equiv A_{\mu}^aJ_a\equiv\omega_{\mu}+\frac{1}{l}e_{\mu}; \qquad 
\bar{A}_{\nu}\equiv\bar{A}_{\nu}^aJ_a\equiv\omega_{\nu}-\frac{1}{l} e_{\nu}, 
\nn\\
\eea
where the SL(2) generator is: 
\bea
&&
J_0\equiv\begin{pmatrix} 
0&-\frac{1}{2}
\\
\frac{1}{2}&0
\end{pmatrix}, \ 
J_1\equiv\begin{pmatrix}
0&\frac{1}{2}
\\
\frac{1}{2}&0
\end{pmatrix}, \ 
J_2\equiv\begin{pmatrix} 
\frac{1}{2}&0
\\ 
0&-\frac{1}{2}
\end{pmatrix}; 
\nn\\ 
&& 
\lbrack J^a, J^b\rbrack=\epsilon^{abc}, \ 
\mathrm{Tr}(J^aJ^b)=\frac{1}{2}\eta^{ab}.
\eea 
The dreibeine indices and Lie algebra are raised and lowered with 
\bea
\eta\equiv\mathrm{diag}(-1,1,1).
\eea 
To identify the gauge formulation by the 3d pure Einstein gravity theory, we define the constant $k$ as
\bea
k\equiv\frac{l}{4G_3},
\eea
where 
\bea
\frac{1}{l^2}\equiv-\Lambda.
\eea
Note that the $G_3$ is the three-dimensional bare gravitational constant. 
\\

\noindent 
The bulk AdS$_3$ solution for the gauge field is: 
\bea
&&
lA
\nn\\
&=&\sqrt{-\Lambda r^2+1}J_0 dx^++\sqrt{-\Lambda} rJ_1dx^++\frac{dr}{\sqrt{-\Lambda r^2+1}}J_2; 
\nn\\
&&
l\bar{A}
\nn\\
&=&-\sqrt{-\Lambda r^2+1}J_0 dx^-+\sqrt{-\Lambda}r J_1dx^--\frac{dr}{\sqrt{-\Lambda r^2+1}}J_2,  
\nn\\
\eea
where 
\bea
x^{\pm}\equiv t\pm\theta. 
\eea
The dreibeine generates the AdS$_3$ metric as that 
\bea
g_{\mu\nu}=2\mathrm{Tr}(e_{\mu}e_{\nu}). 
\eea 
When choosing the unit cosmological constant
 \bea
 \Lambda=-1\,,
 \eea 
the geometry implies the boundary conditions of the gauge fields:
\bea
(E_{\theta}^+A_t-E_t^+A_{\theta})|_{r\rightarrow\infty}&=&0; 
\nn\\
(E_{\theta}^-\bar{A}_t-E_t^-\bar{A}_{\theta})|_{r\rightarrow\infty}&=&0.
\eea
We label the indices of boundary spacetimes as $\tilde{\mu}=t, \theta$. The boundary zweibein provides the boundary metric
\bea
g_{\tilde{\mu}\tilde{\nu}}\equiv\frac{1}{2}(E^+_{\tilde{\mu}}E^-_{\tilde{\nu}}+E^-_{\tilde{\mu}}E^+_{\tilde{\nu}}).  
\eea 
We can integrate over the functional space that is free of singular metrics. 
The same restriction picks out a component of the flat connection to be the product of two copies of Teichmüller space, which is the phase space of gravity \cite{Mikhaylov:2017ngi,Collier:2023fwi}.

%The measure in this gauge formulation is $\int {\cal D}A{\cal D}\bar{A}$, not the same as in the metric formulation \cite{Collier:2023fwi}. 
%The singular metric (no inverse) exists in the gauge formulation but not in the metric formulation. 
%Without the non-perturbative correction, the metric formulation is equivalent to the gauge formulation, and the measure is over Teichmüller space \cite{Collier:2023fwi}. 

\subsection{Boundary Theory}
\noindent
The asymptotic behavior of the gauge fields on the torus is that:
  \bea
A|_{r\rightarrow\infty}=
\begin{pmatrix}
\frac{dr}{2r}& 0
\\
rE^+& -\frac{dr}{2r}
\end{pmatrix};
\qquad
\bar{A}|_{r\rightarrow\infty}=
\begin{pmatrix}
-\frac{dr}{2r}& -rE^{-}
\\
0& \frac{dr}{2r}
\end{pmatrix},
\nn\\
\eea
The $E^+$ and $E^-$ determine the shape of the boundary manifold. 
The parameterization of the SL(2, $\mathbb{R}$) transformations are:
\bea
g_{\mathrm{SL(2)}}&=&
\begin{pmatrix}
1& 0
\\
F& 1
\end{pmatrix}
\begin{pmatrix}
\lambda & 0
\\
0& \frac{1}{\lambda}
\end{pmatrix}
\begin{pmatrix}
1 &\Psi
\\
0& 1
\end{pmatrix}; 
\nn\\
\bar{g}_{\mathrm{SL(2)}}&=&
\begin{pmatrix}
1& -\bar{F}
\\
0& 1
\end{pmatrix}
\begin{pmatrix}
\frac{1}{\bar{\lambda}} & 0
\\
0& \bar{\lambda}
\end{pmatrix}
\begin{pmatrix}
1 &0
\\
-\bar{\Psi}& 1
\end{pmatrix}.
\eea 
Because we consider the positive $\lambda$, the decomposition covers PSL(2, $\mathbb{R}$) but not SL(2, $\mathbb{R}$). 
\\

\noindent 
The asymptotic form of the condition of a flat connection
\bea
g^{-1}_{\mathrm{SL(2)}}\partial_{\theta}g_{\mathrm{SL(2)}}|_{r\rightarrow\infty}
&=&A_{\theta}|_{r\rightarrow\infty}, 
\nn\\ 
\bar{g}^{-1}_{\mathrm{SL(2)}}\partial_{\theta}\bar{g}_{\mathrm{SL(2)}}|_{r\rightarrow\infty}
&=&\bar{A}_{\theta}|_{r\rightarrow\infty}
\eea
leads to the following parameterization at the boundary \cite{Cotler:2018zff}:
\bea
&&
\lambda=\sqrt{\frac{r E_{\theta}^+}{\partial_{\theta}F}}; \ 
\Psi=-\frac{1}{2rE_{\theta}^+}\frac{\partial_{\theta}^2F}{\partial_{\theta}F}, 
\nn\\
&& 
\bar{\lambda}=\sqrt{\frac{r E_{\theta}^-}{\partial_{\theta}\bar{F}}}; \  
\bar{\Psi}=-\frac{1}{2rE_{\theta}^-}\frac{\partial_{\theta}^2\bar{F}}{\partial_{\theta}\bar{F}}. 
\eea
The boundary theory before inserting the constraint is \cite{Cotler:2018zff}
\bea
&&S_{\mathrm{B}}
\nn\\
&=&\frac{k}{\pi}\int dtd\theta\ \bigg(\frac{(\partial_{\theta}\lambda)(D_-\lambda)}{\lambda^2}+\lambda^2(\partial_{\theta}F)(D_-\Psi)\bigg)
\nn\\
&&
-\frac{k}{\pi}\int dtd\theta\ \bigg(\frac{(\partial_{\theta}\bar{\lambda})(D_+\bar{\lambda})}{\bar{\lambda}^2}+\bar{\lambda}^2(\partial_{\theta}\bar{F})(D_+\bar{\Psi})\bigg),
\nn\\
\eea
where
 \bea
D_+\equiv\frac{1}{2} \partial_t-\frac{1}{2}\frac{E_{t}^-}{E_{\theta}^-}\partial_{\theta}; \qquad D_-\equiv\frac{1}{2}\partial_t-\frac{1}{2}\frac{E_{t}^+}{E_{\theta}^+}\partial_{\theta}.
\eea
Here we only integrate out $A_t$ and $\bar{A}_t$ without using the boundary constraint. 
The Weyl transformation now also transforms the boundary field as follows
\bea
E^{\pm}\rightarrow \exp(\sigma)E^{\pm}. 
\eea
After inserting the constraint, the boundary theory on the torus is 2d Schwarzian theory \cite{Cotler:2018zff}, 
 \bea
&&S_{\mathrm{Gb}}
\nn\\
&=&\frac{k}{2\pi}\int dtd\theta\ \bigg(\frac{3}{2}\frac{(D_-\partial_{\theta}{\cal F})(\partial_{\theta}^2{\cal F})}{(\partial_{\theta}{\cal F})^2}-\frac{D_-\partial_{\theta}^2{\cal F}}{\partial_{\theta}{\cal F}}
\bigg)
\nn\\
&&-\frac{k}{2\pi}\int dtd\theta\ \bigg(\frac{3}{2}\frac{(D_+\partial_{\theta}\bar{{\cal F}})(\partial_{\theta}^2\bar{{\cal F}})}{(\partial_{\theta}\bar{{\cal F}})^2}-\frac{D_+\partial_{\theta}^2\bar{{\cal F}}}{\partial_{\theta}\bar{{\cal F}}}
\bigg).
\nn\\
\eea
\\ 

\noindent 
This boundary theory does not preserve the modular symmetry on the torus \cite{Cotler:2018zff}. 
It is easier to understand this point by calculating the torus partition function. 
 The global AdS$_3$ is on the Euclidean manifold. 
 The boundary is a torus.  
 The coordinate of the torus is 
 \bea
 z\equiv\theta+i\psi,  
 \eea
 where $\psi\equiv it$ (Euclidean time).  
 The identification of the torus is: 
 \bea
z\sim z+2\pi\tau, 
 \eea
 where $\tau$ is a complex structure. 
 \\
 
 \noindent
 We transform the boundary field ${\cal F}$ to $\phi$ as 
 \bea
 {\cal F}\equiv\tan\bigg(\frac{\phi}{2}\bigg). 
 \eea
 It is easier to solve the on-shell solution. 
 The saddle-point is unique up to the gauge redundancy \cite{Cotler:2018zff}
 \bea
 \phi_0=\theta-\frac{\mathrm{Re}(\tau)}{\mathrm{Im}(\tau)}\psi. 
 \eea
 Then we obtain the boundary condition for the boundary condition on global AdS$_3$, in which the spatial circle becomes contractible (we call A cycle of the torus) \cite{Cotler:2018zff}: 
 \bea
 &&
 \phi(\psi, \theta+2\pi)=\phi(\psi, \theta)+2\pi;  
 \nn\\
 &&
 \phi\big(\psi+2\pi\mathrm{Im}(\tau), \theta+2\pi\mathrm{Re}(\tau)\big)=\phi(\psi, \theta).  
 \eea
 \\
 
 \noindent 
 For the B cycle (thermal circle or $\psi$ direction becomes contractible), the bulk manifold is the Euclidean BTZ black hole. 
 The on-shell solution is also unique \cite{Cotler:2018zff}
 \bea
 \phi_0=\frac{\psi}{\mathrm{Im}(\tau)}. 
 \eea
 The boundary condition is \cite{Cotler:2018zff}:
\bea
&&
 \phi(\psi, \theta+2\pi)=\phi(\psi, \theta);
 \nn\\
 && 
 \phi\big(\psi+2\pi\mathrm{Im}(\tau), \theta+2\pi\mathrm{Re}(\tau)\big)=\phi(\psi, \theta)+2\pi .  
 \eea
After the modular transformation 
\bea
\tau\rightarrow-\frac{1}{\tau}, 
\eea 
the partition function of the A-cycle becomes the B-cycle one \cite{Cotler:2018zff}. 
Because the partition function is not invariant under the modular transformation
\bea
\tau\rightarrow -\frac{1}{\tau}, 
\eea
this theory does not have a modular symmetry \cite{Cotler:2018zff}. 
Because the boundary theory has the on-shell Lorentz and Weyl symmetries, it is invariant under the conformal transformation. 
The loss of modular symmetry is due to the non-periodic boundary condition. 
The conventional CFT on the torus requires periodic boundary conditions for each direction. 
Hence a conformal but not modular-invariant theory arises at the boundary of pure Einstein gravity theory on asymptotically AdS$_3$ spacetime. 
Because the Lagrangian does not have the covariant form under coordinate transformations, the Weyl anomaly may not come from the Liouville theory. 
Hence we check the Weyl anomaly on the cylinder and torus manifolds.  

\subsection{Cylinder$\rightarrow$Sphere}
\noindent
The spin connection $\omega^a$ satisfies the torsionless condition (or by the variation of $\omega^a$)
\bea
de_a+\epsilon_{abc}\omega^b\wedge e^c=0. 
\eea
The Weyl transformation (from the cylinder manifold to the sphere manifold): 
\bea
e^a\rightarrow\exp\big(\sigma(t)\big)e^a=\sech(\psi)e^a  
\eea
leads the asymptotic boundary condition to the gauge fields: 
\bea
A_{\theta}|_{r\rightarrow\infty}&=&
\begin{pmatrix}
-\frac{1}{2}\partial_t\sigma& 0
\\
re^{\sigma}E_{\theta}^+& \frac{1}{2}\partial_t\sigma 
\end{pmatrix}; 
\nn\\ 
\bar{A}_{\theta}|_{r\rightarrow\infty}&=&
\begin{pmatrix}
-\frac{1}{2}\partial_t\sigma& -re^{\sigma}E_{\theta}^-
\\
0& \frac{1}{2}\partial_t\sigma  
\end{pmatrix}.
\eea
The boundary conditions become:
\bea
(E_{\theta}^+A_t-E_t^+A_{\theta}+E_t^+A_{\theta}^2J_2)|_{r\rightarrow\infty}&=&0; 
\nn\\ 
(E_{\theta}^-\bar{A}_t-E_t^-\bar{A}_{\theta}+E_t^-\bar{A}_{\theta}^2J_2)|_{r\rightarrow\infty}&=&0, 
\eea
where 
\bea
A_{\theta}^2=\bar{A}_{\theta}^2=-\partial_t\sigma. 
\eea
The variation of $A_{\theta}$ and $\bar{A}_{\theta}$ shows the non-vanishing boundary term
\bea
-\frac{k}{8\pi}\int dt d\theta\ \frac{E_t^+}{E_{\theta}^+}\delta(A_{\theta}^2A_{\theta}^2)
+\frac{k}{8\pi}\int dt d\theta\ \frac{E_t^-}{E_{\theta}^-}\delta(\bar{A}_{\theta}^2\bar{A}_{\theta}^2).
\nn\\
\eea
Therefore, we need to introduce the additional boundary term 
\bea
S_{\mathrm{B}1}=\frac{k}{8\pi}\int dt d\theta\ \frac{E_t^+}{E_{\theta}^+}A_{\theta}^2A_{\theta}^2
-\frac{k}{8\pi}\int dt d\theta\ \frac{E_t^-}{E_{\theta}^-}\bar{A}_{\theta}^2\bar{A}_{\theta}^2. 
\nn\\
\eea
The cylinder manifold corresponds to that:  
\bea
E_t^+=1=-E_t^-=E_{\theta}^+=E_{\theta}^-. 
\eea
Hence we obtain that 
\bea
S_{\mathrm{B}1}
=\frac{k}{4\pi}\int dtd\theta\ (\partial_t\sigma)(\partial_t\sigma). 
\eea
%When $\sigma$ is a constant, the gauge fields reduce to the torus case. 
The boundary constraints after the Weyl transformation become: 
\bea
&&
\lambda=e^{\frac{\sigma}{2}}\lambda^+, \ 
\Psi=-\frac{e^{-\sigma}}{2E_{\theta}^+r}\big(\partial_{\theta}(\ln{\cal F})-\partial_t\sigma\big), \ 
{\cal F}\equiv e^{-\sigma}{\cal F}^+;  
\nn\\
&&
\bar{\lambda}=e^{\frac{\sigma}{2}}\lambda^-, \ 
\bar{\Psi}=-\frac{e^{-\sigma}}{2E_{\theta}^-r}\big(\partial_{\theta}(\ln\bar{{\cal F}})+\partial_t\sigma\big), \ 
\bar{{\cal F}}\equiv e^{-\sigma}{\cal F}^-, 
\nn\\
\eea 
where 
\bea
&&
\lambda^{+}\equiv\sqrt{\frac{rE_{\theta}^+}{\partial_{\theta}F}}, \ 
\lambda^{-}\equiv\sqrt{\frac{rE_{\theta}^-}{\partial_{\theta}\bar{F}}}; 
\nn\\
&& 
{\cal F}^+\equiv\frac{\partial_{\theta}F}{E_{\theta}^+}, \ 
{\cal F}^-\equiv\frac{\partial_{\theta}\bar{F}}{E_{\theta}^-}. 
\eea
\\

\noindent 
The Weyl transformation induces the additional terms from the cylinder case 
\bea
&&
\delta S_{\mathrm{B}}
\nn\\
&=&\frac{k}{\pi}\int dt d\theta\ \bigg(
-\frac{1}{4}\frac{(\partial_{\theta}\lambda^+)(\partial_t\sigma)}{\lambda}
+\frac{1}{4}\frac{(\partial_{\theta}\lambda^-)(\partial_t\sigma)}{\bar{\lambda}}
\nn\\
&&
-\frac{1}{2}(\partial_t\sigma)(\partial_t\sigma)
+\frac{1}{2}\partial_t^2\sigma
\bigg).  
\eea
The coupling term between the background field $\sigma$ and $\lambda^+, \lambda^-$ vanishes, up to a total derivative term (for the $\theta$). 
Therefore, the background $\sigma$ is the only source of the Weyl anomaly 
\bea
\delta S_{\mathrm{B}}
=\frac{k}{\pi}\int dt d\theta\ \bigg(
-\frac{1}{2}(\partial_t\sigma)(\partial_t\sigma)
+\frac{1}{2}\partial_t^2\sigma
\bigg). 
\eea
We combine all terms relevant to the anomaly: 
\bea
&&
\delta S_{\mathrm{B}}+S_{\mathrm{B}1}
\nn\\
&=&\frac{k}{\pi}\int dt d\theta\ \bigg(
-\frac{1}{4}(\partial_t\sigma)(\partial_t\sigma)
+\frac{1}{2}\partial_t^2\sigma
\bigg)
. 
\eea 
The first term is the Liouville theory's Weyl anomaly 
\bea
-\frac{k}{4\pi}\int dt d\theta\ (\partial_t\sigma)(\partial_t\sigma). 
\eea
The boundary term is necessary for obtaining the Liouville theory. 
Therefore, the boundary term on different manifolds is different, not the same as the claim of Ref. \cite{Cotler:2018zff}. 
Now we check the second term from the specific transformation $\exp(\sigma)=\sech(\psi)$. 
The Weyl anomaly for the Euclidean time coordinate is 
\bea
\frac{k}{\pi}\int d\psi d\theta\ \bigg(
-\frac{1}{4}(\partial_{\psi}\sigma)(\partial_{\psi}\sigma)
+\frac{1}{2}\partial_{\psi}^2\sigma
\bigg)
,  
\eea
where 
\bea
-\infty<\psi<\infty, \ 
0<\theta\le 2\pi. 
\eea
Therefore, we can show that the second term (or one-dimensional boundary term) is finite
\bea
\frac{k}{2\pi}\int_{-\infty}^{\infty} d\psi\int_0^{2\pi} d\theta\ \partial^2_{\psi}\sigma
=-2k. 
\eea
When computing the divergent quantity, one can ignore the finite term. 
Therefore, it is applicable for the CFT computation on a sphere manifold to compute the entanglement entropy of a single interval. 

\subsection{General Weyl Transformation on Torus Manifold}
\noindent
Now we consider a general Weyl transformation on the torus. 
The transformation now depends on $t$ and $\theta$. 
The spin connection becomes
\bea
\omega^2=-(\partial_{\theta}\sigma)dt-(\partial_t\sigma)d\theta. 
\eea
Therefore, the asymptotic boundary condition for $A_{\theta}$ and $\bar{A}_{\theta}$ remains the same. 
The constraints on the boundary are also the same. 
Now the boundary condition is different:
\bea
&&
(E_{\theta}^+A_t-E_t^+A_{\theta}-E_{\theta}^+A_t^2J_2+E_t^+A_{\theta}^2J_2)|_{r\rightarrow\infty}=0; 
\nn\\
&& 
(E_{\theta}^-\bar{A}_t-E_t^-\bar{A}_{\theta}-E_{\theta}^-\bar{A}_t^2+E_t^-\bar{A}_{\theta}^2J_2)|_{r\rightarrow\infty}=0, 
\eea
where 
\bea
A_{\theta}^2=\bar{A}_{\theta}^2=-\partial_t\sigma; \ 
A_t^2=\bar{A}_t^2=-\partial_{\theta}\sigma.  
\eea
\\

\noindent
The variation of $A_{\theta}$ and $\bar{A}_{\theta}$ shows the non-vanishing boundary term
\bea
&&
-\frac{k}{8\pi}\int dt d\theta\ \bigg(\frac{E_t^+}{E_{\theta}^+}\delta(A_{\theta}^2A_{\theta}^2)
-\delta(A_{t}^2A_{t}^2)\bigg)
\nn\\
&&
+\frac{k}{8\pi}\int dt d\theta\ \bigg(\frac{E_t^-}{E_{\theta}^-}\delta(\bar{A}_{\theta}^2\bar{A}_{\theta}^2)
-\delta(\bar{A}_{t}^2\bar{A}_{t}^2)\bigg). 
\eea 
Therefore, we need to introduce the additional boundary term, which produces the same anomaly term even from a more general Weyl transformation: 
\bea
&&
S_{\mathrm{B}1}
\nn\\
&=&\frac{k}{8\pi}\int dt d\theta\ \frac{E_t^+}{E_{\theta}^+}A_{\theta}^2A_{\theta}^2
-\frac{k}{8\pi}\int dtd\theta\ A_t^2A_t^2
\nn\\
&&
-\frac{k}{8\pi}\int dt d\theta\ \frac{E_t^-}{E_{\theta}^-}\bar{A}_{\theta}^2\bar{A}_{\theta}^2
+\frac{k}{8\pi}\int dtd\theta\ \bar{A}_t^2\bar{A}_t^2
\nn\\
&=&\frac{k}{8\pi}\int dt d\theta\ \frac{E_t^+}{E_{\theta}^+}A_{\theta}^2A_{\theta}^2
-\frac{k}{8\pi}\int dt d\theta\ \frac{E_t^-}{E_{\theta}^-}\bar{A}_{\theta}^2\bar{A}_{\theta}^2
\nn\\
&=&\frac{k}{4\pi}\int dtd\theta\ (\partial_t\sigma)(\partial_t\sigma)
. 
\eea
The torus manifold corresponds to that:  
\bea
E_t^+=1=-E_t^-=E_{\theta}^+=E_{\theta}^-. 
\eea
The boundary constraints after the Weyl transformation become: 
\bea
\lambda&=&e^{\frac{\sigma}{2}}\lambda^+,
\nn\\ 
\Psi&=&-\frac{e^{-\sigma}}{2E_{\theta}^+r}\big(\partial_{\theta}(\ln{\cal F}^+)-\partial_{\theta}\sigma-\partial_t\sigma\big), 
\nn\\ 
{\cal F}&\equiv& e^{-\sigma}{\cal F}^+;  
\nn\\
\bar{\lambda}&=&e^{\frac{\sigma}{2}}\lambda^-, 
\nn\\ 
\bar{\Psi}&=&-\frac{e^{-\sigma}}{2E_{\theta}^-r}\big(\partial_{\theta}(\ln{\cal F}^-)-\partial_{\theta}\sigma+\partial_t\sigma\big), 
\nn\\ 
\bar{{\cal F}}&\equiv& e^{-\sigma}{\cal F}^-.
\eea 
Because we consider the torus manifold, we can do the integration by part for each direction. 
Therefore, the Weyl transformation generates the additional term that 
\bea
\delta S_{\mathrm{B}}=\frac{k}{\pi}\int dtd\theta\ \bigg(
-\frac{1}{2}(\partial_t\sigma)(\partial_t\sigma)
+\frac{1}{4}(\partial_{\theta}\sigma)(\partial_{\theta}\sigma)
\bigg). 
\nn\\
\eea 
The coupling term between the background and dynamical fields vanishes through the integration by part: 
\bea
&&
\frac{k}{\pi}\int dt d\theta\ \bigg(\frac{1}{4}(\partial_{\theta}\sigma)(\partial_t\ln\lambda^+)
+\frac{1}{4}(\partial_{t}\sigma)(\partial_{\theta}\ln\lambda^+)
\nn\\
&&
-\frac{1}{2}(\partial_{\theta}\sigma)(\partial_{\theta}\ln\lambda^+)
\nn\\
&&
+\frac{1}{4}(\partial_t\sigma)(\partial_{\theta}\ln{\cal F}^+)-\frac{1}{4}(\partial_{\theta}\sigma)(\partial_{\theta}\ln{\cal F}^+)
\nn\\
&&
-\frac{1}{4}(\partial_{\theta}\sigma)(\partial_t\ln\lambda^-)
-\frac{1}{4}(\partial_{t}\sigma)(\partial_{\theta}\ln\lambda^-)
\nn\\
&&
-\frac{1}{2}(\partial_{\theta}\sigma)(\partial_{\theta}\ln\lambda^-)
\nn\\
&&
-\frac{1}{4}(\partial_t\sigma)(\partial_{\theta}\ln{\cal F}^-)-\frac{1}{4}(\partial_{\theta}\sigma)(\partial_{\theta}\ln{\cal F}^-)
\bigg)
\nn\\
&=&\frac{k}{\pi}\int dt d\theta\ \bigg(\frac{1}{2}(\partial_{\theta}\sigma)(\partial_t\ln\lambda^+)
-\frac{1}{2}(\partial_{\theta}\sigma)(\partial_{\theta}\ln\lambda^+)
\nn\\
&&
+\frac{1}{4}(\partial_t\sigma)(\partial_{\theta}\ln{\cal F}^+)-\frac{1}{4}(\partial_{\theta}\sigma)(\partial_{\theta}\ln{\cal F}^+)
\nn\\
&&
-\frac{1}{2}(\partial_{\theta}\sigma)(\partial_t\ln\lambda^-)
-\frac{1}{2}(\partial_{\theta}\sigma)(\partial_{\theta}\ln\lambda^-)
\nn\\
&&
-\frac{1}{4}(\partial_t\sigma)(\partial_{\theta}\ln{\cal F}^-)-\frac{1}{4}(\partial_{\theta}\sigma)(\partial_{\theta}\ln{\cal F}^-)
\bigg)
\nn\\
&=&\int dtd\theta\ \bigg\lbrack\frac{1}{4}(\partial_t\sigma)\bigg(\partial_{\theta}\ln\big({\cal F}^+(\lambda^+)^2\big)\bigg)
\nn\\
&&
-\frac{1}{4}(\partial_{\theta}\sigma)\bigg(\partial_{\theta}\ln\big({\cal F}^+(\lambda^+)^2\big)\bigg)
\nn\\
&&
-\frac{1}{4}(\partial_t\sigma)\bigg(\partial_{\theta}\ln\big({\cal F}^-(\lambda^-)^2\big)\bigg)
\nn\\
&&
-\frac{1}{4}(\partial_{\theta}\sigma)\bigg(\partial_{\theta}\ln\big({\cal F}^-(\lambda^-)^2\big)\bigg)
\bigg\rbrack
\nn\\
&=&
0.  
\eea
We use: 
\bea
{\cal F}^+(\lambda^+)^2={\cal F}^-(\lambda^-)^2=r
\eea
in the last equality. 
Therefore, the Weyl anomaly induces the Liouville theory 
\bea
&&
\delta S_{\mathrm{B}}+S_{\mathrm{B}1}
\nn\\
&=&\frac{k}{\pi}\int dtd\theta\ \bigg(-\frac{1}{4}(\partial_t\sigma)(\partial_t\sigma)
+\frac{1}{4}(\partial_{\theta}\sigma)(\partial_{\theta}\sigma)\bigg). 
\eea
Hence we can apply the conventional CFT result to compute the mutual information for disjoint intervals. 

\section{Modular Averaging}
\label{sec:3} 
\noindent
The sum over all geometries with the asymptotic AdS$_3$ boundary condition is equivalent to summing over PSL(2, $\mathbb{Z}$)/$\mathbb{Z}$ group (or modular group) in the boundary theory \cite{Maloney:2007ud,Manschot:2007ha}. Modular averaging includes the non-perturbative contribution. 
Because the torus partition function is one-loop exact \cite{Cotler:2018zff}, the averaging also has the analytical expression. 
Comparing the result with and without averaging helps examine the phase transition from the analytical solution.   
\\

\noindent 
We first define the group SL(2, $\mathbb{Z}$) as 2 by 2 matrices over $\mathbb{Z}$ having determinant 1. 
In other words, the elements $a_1, b_1, c_1, d_1$ form a 2 by 2 matrix
\bea
\begin{pmatrix}
a_1 & b_1
\\
c_1 & d_1
\end{pmatrix}
\eea 
with 
\bea
a_1d_1-b_1c_1=1, \ a_1, b_1, c_1, d_1\in \mathbb{Z}.
\eea 
\\

\noindent 
Because the partition function only depends on the variable $(a_1\tau+b_1)/(c_1\tau+d_1)$, we only sum over the relative primes $c_1$ and $d_1$ whose greatest common divisor $(c_1, d_1)$ is one,
\bea
(c_1, d_1)=1\,.
\eea
If we have two solutions for $a_1, b_1$ and $a_2, b_2$, they satisfy the following equation
\bea
(a_1-a_2)d_1=(b_1-b_2)c_1. 
\eea
Therefore, we can show that: 
\bea
a_1-a_2=mc_1, \ b_1-b_2=md_1, 
\eea
where $m$ is an arbitrary integer. 
When we choose the solution as that: 
\bea
a_1\rightarrow a_1+mc_1; \ b\rightarrow b_1+md_1, 
\eea
it implies 
\bea
\frac{a_1\tau+b_1}{c_1\tau+d_1}\rightarrow\frac{a_1\tau+b_1}{c_1\tau+d_1}+m.
\eea
Hence we can find that:
\bea
q\rightarrow qe^{2\pi im}=q, 
\eea
where 
\bea
q=e^{2\pi i\tau}. 
\eea
Because the partition function only depends on $q$, different choices of $m$ correspond to the same contribution. 
Therefore, we choose $m=0$ for the summation (or do not sum over $a_1$ and $b_1$). 
\\

\noindent 
The partition function of 2d Schwarzian theory on a torus is \cite{Cotler:2018zff}
\bea
Z_T(\tau)=|q|^{-\frac{c}{12}}\frac{1}{\prod^{\infty}_{n=2}|1-q^n|^2}.
\eea
The central charge $c$ is \cite{Huang:2019nfm,Huang:2020tjl}
\bea
c=6k+13.   
\eea 
The shifting of central charge or 13 is due to the one-loop correction of the boundary theory \cite{Cotler:2018zff}. 
Therefore, the $1/c$ is proportional to the physical gravitational constant. 
Because the partition function is one-loop exact \cite{Cotler:2018zff}, the relation of central charge is not an approximation.
\\

\noindent
By the Dedekind $\eta$ function, 
\bea
\eta(\tau)=q^{\frac{1}{24}}\prod^{\infty}_{n=1}(1-q^n), 
\eea
we rewrite the partition function as that
\bea
Z_T(\tau)=\frac{1}{|\eta(\tau)|^2}|q|^{-\frac{1}{12}(c-1)}|1-q|^2. 
\eea
The partition function becomes the manifestly modular invariant form \cite{Maloney:2007ud}
\bea
Z_M(\tau)=\sum_{c_1, d_1;\ (c_1, d_1)=1}Z_T\bigg(\frac{a_1\tau+b_1}{c_1\tau+d_1}\bigg), 
\eea
where $a_1, b_1, c_1, d_1$ are the elements of SL(2, $\mathbb{Z}$). 
Because $\sqrt{\mathrm{Im}(\tau)}|\eta(\tau)|^2$ is the modular invariant, we can rewrite the partition function as that
\bea
&&
Z_M(\tau)
\nn\\
&=&\frac{1}{\sqrt{\mathrm{Im}(\tau)}|\eta(\tau)|^2}
\sum_{c_1, d_1}(\sqrt{\mathrm{Im}(\tau)}|q|^{-\frac{1}{12}(c-1)}|1-q|^2)|_{M}, 
\nn\\
\eea
where $(\cdots)|_M$ means that 
\bea
\tau\rightarrow\frac{a_1\tau+b_1}{c_1\tau+d_1} 
\eea 
when $\mathrm{Im}(\tau)>0$. 
Later we will expand $|1-q|^2$ and express the $Z(\tau)$ as a sum of the Poincaré series \cite{Maloney:2007ud}, defined as 
\bea
P(\tau; n, m)\equiv\sum_{c_1, d_1; (c_1, d_1)=1}(\sqrt{\mathrm{Im}\tau}q^{-n}\bar{q}^{-m})|_M.
\eea
The partition function becomes 
\bea
&&
Z_M(\tau)
\nn\\
&=&\frac{1}{\sqrt{\mathrm{Im}(\tau)}|\eta(\tau)|^2}
\nn\\
&&\times\bigg\lbrack
P\bigg(\tau; \frac{c-1}{24}, \frac{c-1}{24}\bigg)
-P\bigg(\tau; \frac{c-1}{24}-1, \frac{c-1}{24}\bigg)
\nn\\
&&
-P\bigg(\tau; \frac{c-1}{24}, \frac{c-1}{24}-1\bigg)
\nn\\
&&
+P\bigg(\tau; \frac{c-1}{24}-1, \frac{c-1}{24}-1\bigg)\bigg\rbrack. 
\eea

\section{Rényi-2}
\label{sec:4}
\noindent 
We calculate the Rényi-2 mutual information for two disjoint intervals. 
The end points of the first interval are $u_1, v_1$ ($v_1>u_1$) and the second interval are $u_2, v_2$ ($v_2>u_2$). 
The cross ratio 
\bea
z=\frac{(v_1-u_1)(v_2-u_2)}{(u_2-u_1)(v_2-v_1)}
\eea
is relevant to the complex structure $\tau$ as in the following: 
\bea
z=\frac{\theta_2^4(\tau)}{\theta_3^4(\tau)}, \qquad 
1-z=\frac{\theta_4^4(\tau)}{\theta_3^4(\tau)}=\frac{\theta_2^4\big(-\frac{1}{\tau}\big)}{\theta_3^4\big(-\frac{1}{\tau}\big)}.  
\eea
The notation of theta functions is:
\bea
\theta_2(\tau)&\equiv&\sum_{j=-\infty}^{\infty}q^{\frac{1}{2}\big(j+\frac{1}{2}\big)^2};
\nn\\
\theta_3(\tau)&\equiv&\sum_{j=-\infty}^{\infty}q^{\frac{j^2}{2}};
\nn\\
\theta_4(\tau)&\equiv&\sum_{j=-\infty}^{\infty}(-1)^jq^{\frac{j^2}{2}},  
\eea
where 
\bea
q\equiv e^{2\pi i\tau}. 
\eea
Now the complex structure is an imaginary number 
\bea
\tau=il, \ l\in\mathbb{R}. 
\eea
As $z$ goes from 0 to 1, $l$ goes from $\infty$ to 0. 
When $z$ approaches 0 (or $l\rightarrow\infty$), the separation of two intervals approaches infinity.  
The Rényi-2 mutual information has the expected decay for a long separation of two intervals.    
Our result also shows that the non-perturbative correction kills the first-order phase transition (for the separation of two intervals). 

\subsection{Rényi-2 Mutual Information} 
\noindent
The Rényi entropy of order $q$ is 
\bea
R^{(q)}\equiv\frac{\ln\mathrm{Tr}\rho^q}{1-q}, 
\eea
where $\rho$ is a reduced density matrix with the unit normalization $\mathrm{Tr}(\rho)=1$. 
The difficulty of computing $R^{(q)}$ in Quantum Field Theory (QFT) is that we cannot define a reduced density matrix $\rho$. 
The replica trick provides the path integral over the $q$-sheet space to $\mathrm{Tr}\rho^q$. 
Therefore, we can compute the $q$-sheet partition function, $Z^{(q)}$, to obtain $R^{(q)}$ as that 
\bea
R^{(q)}=\frac{\ln Z^{(q)}-q\ln Z^{(1)}}{1-q}, 
\eea 
where $Z^{(1)}$ is for normalization. 
The Rényi-2 entropy is: 
\bea
R^{(2)}=-\ln\mathrm{Tr}\rho^2=-\ln\frac{Z^{(2)}}{\big(Z^{(1)}\big)^2}.  
\eea
The $q$-sheet partition function for disjoint $n$-intervals is relevant to a surface with $(n-1)(q-1)$ genus \cite{Faulkner:2013yia}. 
Therefore, we will substitute the partition function on a torus (1 genus) to $Z^{(2)}$. 
The $Z^{(1)}$ is only for normalization. 
Therefore, ignoring this term does not affect the qualitative behavior of $R^{(2)}$. 
\\

\noindent 
We introduce the regularized twist operators, $\sigma_1^{\epsilon}, \sigma_{-1}^{\epsilon}$, where $\epsilon$ is a UV cutoff, with scaling dimension
\bea
d_{\sigma}=\frac{c}{12}\bigg(q-\frac{1}{q}\bigg)
\eea
to compute $Z^{(2)}$. 
The twist operators introduce the $q$-fold branch cuts producing the $q$-sheet boundary condition. 
We can read the two-point regularized twist operators from Rényi entropy:
\bea
&&
R^{(q)}_{\lbrack u, v\rbrack}
\nn\\
&=&\frac{1}{1-q}\ln\langle\sigma^{\epsilon}_1(u)\sigma^{\epsilon}_{-1}(v)\rangle
\sim\frac{c}{6}\bigg(1+\frac{1}{q}\bigg)\ln\bigg(\frac{v-u}{\epsilon}\bigg), 
\nn\\
\eea
where $\sim$ is up to the regularization dependent terms. 
Therefore, we obtain 
\bea
\langle\sigma_1^{\epsilon}(u)\sigma_{-1}^{\epsilon}(v)\rangle\sim\bigg(\frac{v-u}{\epsilon}\bigg)^{-\frac{c}{6}\big(n-\frac{1}{n}\big)}.
\eea 
We can choose 
\bea
(u_1, v_1, u_2, v_2)=(0, z, 1, \infty)
\eea
by a conformal transformation. 
When $q=2$, we have a unique twist operator: 
\bea
\sigma\equiv\sigma_1=\sigma_{-1},  
\eea
where the normalized twist operators are 
\bea
\sigma_{\pm 1}\equiv\frac{\sigma_{\pm 1}^{\epsilon}}{\sqrt{\langle\sigma_1^{\epsilon}(0)\sigma_{-1}^{\epsilon}(1)\rangle}}.
\eea
The partition function $Z^{(2)}$ is given by the four-point function \cite{Lunin:2000yv}: 
\bea
Z^{(2)}
=
\langle\sigma(0)\sigma(z)\sigma(1)\tilde{\sigma}(\infty)\rangle
=
\big(2^8z(1-z)\big)^{-\frac{c}{12}}Z_T(\tau), 
\nn\\
\eea
where 
\bea
\tilde{\sigma}_{-1}(\infty)\equiv\lim_{x\rightarrow\infty}x^{2d_{\sigma}}\sigma_{-1}(x). 
\eea
\\

\noindent
The Rényi mutual information of order $q$ is:  
\bea
&&
I^{(q)}_{\lbrack u_1, v_1\rbrack, \lbrack u_2, v_2\rbrack}
\nn\\
&=&R^{(q)}_{\lbrack u_1, v_1 \rbrack}
+R^{(q)}_{\lbrack u_2, v_2 \rbrack}
-R^{(q)}_{\lbrack u_1, v_1 \rbrack\cup\lbrack u_2, v_2\rbrack}
\nn\\
&=&
\frac{1}{q-1}\ln\bigg(\frac{\langle\sigma_1(u_1)\sigma_{-1}(v_1)\sigma_1(u_2)\sigma_{-1}(v_2)\rangle}
{\langle\sigma_1(u_1)\sigma_{-1}(v_1)\rangle\langle\sigma_1(u_2)\sigma_{-1}(v_2)\rangle}
\bigg). 
\eea
Therefore, the Rényi mutual information of order $q$ becomes: 
\bea
&&
I^{(q)}_{\lbrack u_1, v_1\rbrack, \lbrack u_2, v_2\rbrack
}
\nn\\
&=&I^{(q)}_{\lbrack 0, z\rbrack, \lbrack 1, \infty\rbrack}
\nn\\
&\equiv& I^{(q)}(z)
\nn\\
&=&\frac{1}{q-1}\ln\big(z^{2d_{\sigma}}\langle\sigma_1(0)\sigma_{-1}(z)\sigma_1(1)\tilde{\sigma}_{-1}(\infty)\rangle\big). 
\eea
Therefore, the Rényi-2 mutual information is \cite{Headrick:2010zt}: 
\bea
I^{(2)}&=&\ln Z_T(\tau)-\frac{c}{12}\ln\bigg(\frac{2^8(1-z)}{z^2}\bigg)
\nn\\
&=&\ln Z_T(\tau)+c\ln\bigg(\frac{\theta_2(\tau)}{2\eta(\tau)}\bigg), 
\eea
where 
\bea
2\eta^3(\tau)\equiv \theta_2(\tau)\theta_3(\tau)\theta_4(\tau). 
\eea
%The Weyl anomaly only has a finite difference from the Liouville theory. 
%The finite term gives a constant term to the mutual information independent of the subregion. 
%The mutual information decays to zero when the separation of two intervals approaches infinity. 
%Hence the constant term is zero. 
One can apply the result to the AdS$_3$ Einstein gravity theory or sum all bulk geometries with asymptotic AdS$_3$ boundary conditions (replaces $Z_T$ with $Z_M$). 

\subsection{Discussion of Phase Transition}
\begin{figure}
\begin{center}
    \includegraphics[width=0.5\textwidth]{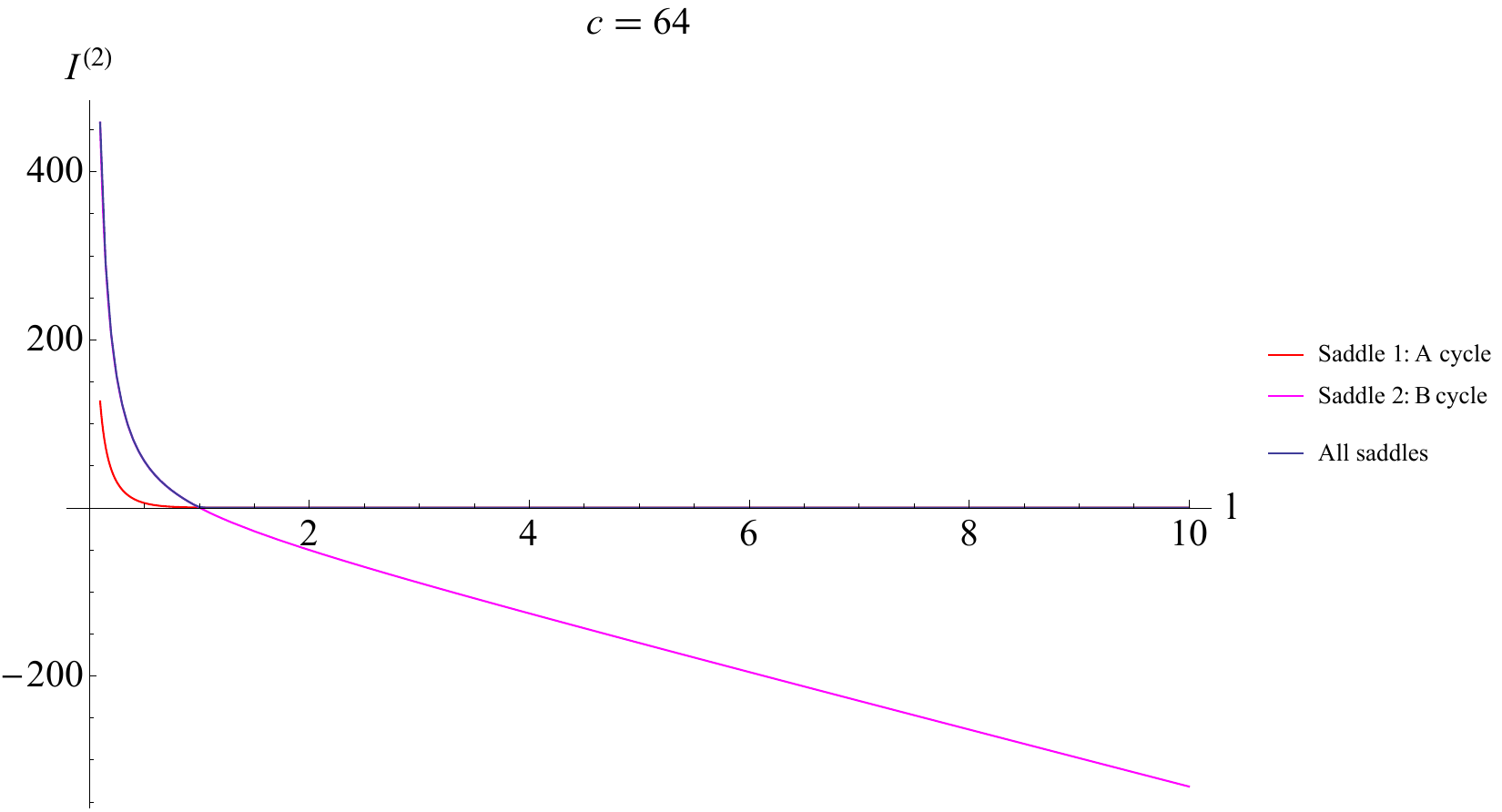}
\end{center}
\caption{We show the Rényi-2 mutual information from summing all saddle points, A-cycle, and B-cycle with $c=64$. 
Not all saddle points decay to zero when $l\rightarrow\infty$ (like B-cycle). 
Therefore, the decay result by summing over all saddle-points is non-trivial. }
\label{I2}
\end{figure}

\begin{figure}
\begin{center}
    \includegraphics[width=0.5\textwidth]{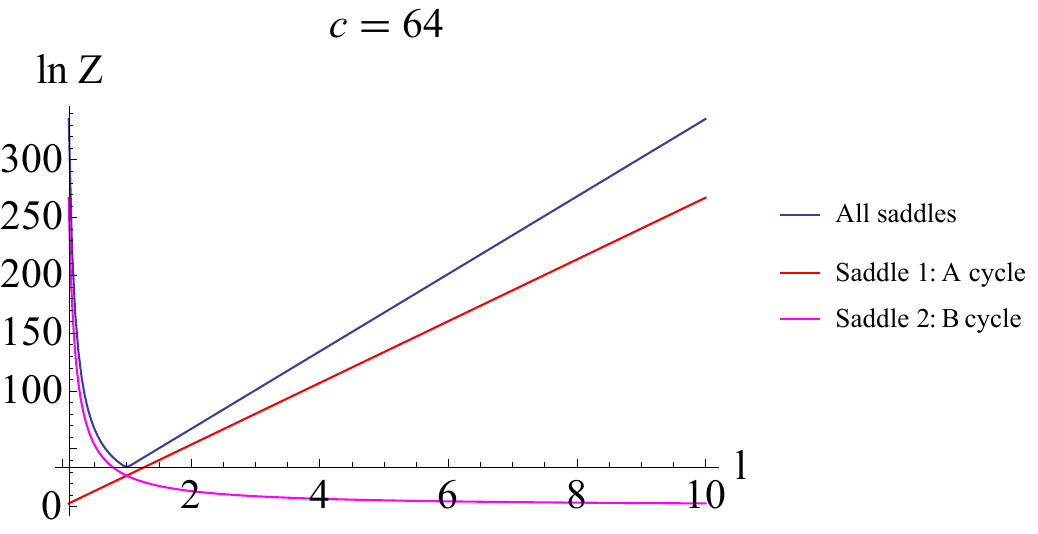}
\end{center}
\caption{The logarithmic partition function is continuous for $l$.
The logarithmic partition function has the transition point between the A-cycle and B-cycle in this plot. 
Therefore, the first-order phase transition can happen when only considering the dominant saddle point.  
}
\label{logZ}
\end{figure}

\begin{figure}
\begin{center}
    \includegraphics[width=0.5\textwidth]{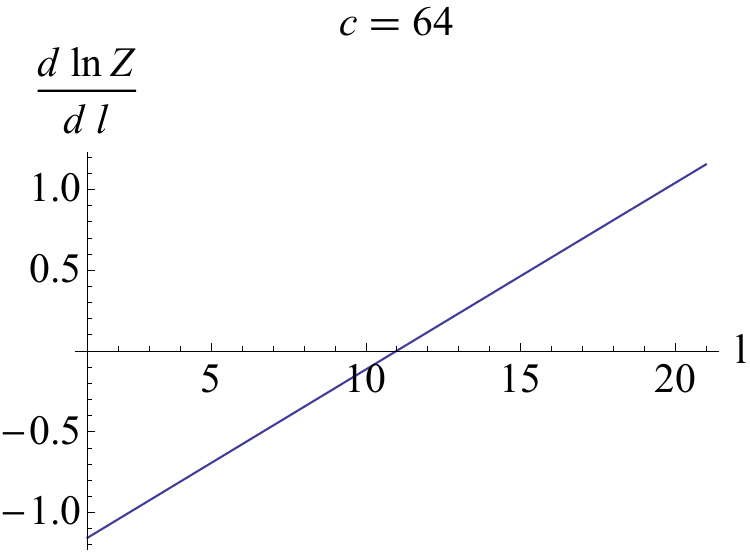}
\end{center}
\caption{The first-order derivative of the logarithmic partition function is a continuous function for $l$. }
\label{derivative}
\end{figure}

\noindent 
We first show the Rényi-2 mutual information $I^{(2)}$ for $c=64$ in Fig. \ref{I2}. 
The mutual information decays to zero for $l\rightarrow\infty$ when implementing a modular average (although the B-cycle solution does not). 
The $I^{(2)}$ now is a smooth function of $l$. 
Therefore, we do not observe the phase transition for $l$, not the same as in only choosing the dominant saddle point. 
The possible source of the phase transition is only the partition function. 
We still see the continuous function for the partition function and its first-order derivative for $l$ as in Figs. \ref{logZ}, \ref{derivative}. 
If we only select the dominant saddle-point, we can observe the transition point in Fig. \ref{logZ}. 
This conclusion still holds without considering the one-loop shift for the central charge. 
Hence the phase transition disappears in our result due to the non-perturbative effect generated by the modular average.
\\

\noindent 
Finally, we comment on the result of Refs. \cite{Huang:2019nfm,Huang:2020tjl}. 
The $I^{(2)}$ decays to zero when the separation of two intervals is infinite. 
The decaying behavior implies the one-loop shift of central charge is 13 for the entanglement entropy of a single interval (same as in the torus partition function). 
The Refs. \cite{Huang:2019nfm,Huang:2020tjl} showed the 26 for the shift. 
The authors used the longer torus ($\tau$ is two times than the usual torus). 
After using the usual torus, the shift becomes 13, consistent with this paper's result.  

\section{Discussion and Conclusion}
\label{sec:5}
\noindent 
It is fascinating to see how the study of the Weyl anomaly and modular average sheds light on the properties of the boundary theory of AdS$_3$ Einstein gravity. 
The fact that the boundary fields do not satisfy the same boundary condition for the A-cycle and B-cycle is an interesting feature that leads to the loss of modular symmetry on the torus \cite{Cotler:2018zff}. 
The Lagrangian formulation is not covariant \cite{Cotler:2018zff}. 
Because this theory is not a conventional CFT, we examine the Weyl anomaly. 
Our result agrees with the Liouville theory for a torus manifold.  
Therefore, we can apply the CFT techniques to calculate the Rényi-2 mutual information. 
The use of CFT techniques to calculate the Rényi-2 mutual information \cite{Headrick:2010zt} and the examination of the first-order phase transition for the separation of two intervals highlight the non-perturbative effects introduced by the averaging over a modular group.
\\

\noindent 
The study of the general Weyl transformation on a torus manifold provides insights into the nature of the Weyl anomaly and its relation to the Liouville theory. 
While the Liouville theory is the answer for a torus manifold, further investigation is needed to determine whether the finite term (1d boundary term) is beyond the Liouville theory, especially in the case of an open boundary.
\\ 

\noindent 
The simple implementation for summing all bulk solutions with asymptotic AdS$_3$ boundary condition after the ensemble average is an intriguing result, although the negative density of state and the loss of unitarity indicate that the boundary theory is not a unitary CFT \cite{Maloney:2007ud}. 
The graviton needs to couple to the matter in our world. 
The loss of unitary for the pure gravity's boundary theory is not too strange. 
The study of pure gravity still has a non-trivial course about the exact bulk/boundary correspondence. 
The fact that the central charge is not proportional to the inverse of the bare gravitational constant \cite{Cotler:2018zff} is another interesting feature that requires further investigation. 
The inverse of the central charge order contains an infinite number of terms in the perturbation of the bare gravitational constant. 
Hence the result of the boundary theory is equivalent to resuming the perturbative gravity result. 
Overall, the study of the Weyl anomaly and modular average provides valuable insights into the properties of the boundary theory of AdS$_3$ Einstein gravity and its relation to CFT techniques and non-perturbative gravity theory. 
The use of modular averaging for resummation could be a promising approach to explore the non-perturbative aspects of gravity theory further.

\section*{Acknowledgments}
\noindent
We thank Yikun Jiang and Mao Tian Tan for their discussion. 
Chen-Te Ma thanks Nan-Peng Ma for his encouragement. 
XH acknowledges the support of the NSFC (Grant No. 12047502 and No. 12247103). 
CTM acknowledges the Nuclear Physics Quantum Horizons program through the Early Career Award (Grant No. DE-SC0021892); 
the YST Program of the APCTP;  
China Postdoctoral Science Foundation, Postdoctoral General Funding: Second Class (Grant No. 2019M652926); 
Foreign Young Talents Program (Grant No. QN20200230017). 
Discussions during the workshops, 
``Recent developments in the holographic principle'', 
``KIAS-YITP 2021 String Theory and Quantum Gravity'', 
``21st String Phenomenology 2022'', 
``East Asia Joint Workshop on Fields and Strings'' , ``15-th International Workshop: Lie Theory and Its Applications in Physics'', 
``28th International Symposium on Particles, Strings and Cosmology (PASCOS 2023)'',  
``Workshop on Cosmology and Quantum Space Time (CQUeST 2023)'', and 
``YITP long-term workshop "Gravity and Cosmology 2024"'', 
were helpful for this work.  

%\appendix

\end{document}